\documentclass{aa}
\usepackage{natbib}
\usepackage[colorlinks=true,citecolor=blue]{hyperref}
\usepackage{soul}
\usepackage{xcolor}
\usepackage[flushleft,online]{threeparttable}
\usepackage{rotating}
\usepackage{pdflscape}
\usepackage{lastpage}
\usepackage[page]{appendix}
\usepackage{upgreek}
\bibpunct{(}{)}{;}{a}{}{,}
\usepackage{graphicx}
\usepackage{txfonts}
\newcommand{\micron}[1]{\,\textmu m}

\begin{document} 

\title{Optimising reference library selection for reference-star differential imaging of discs with SPHERE/IRDIS}

\author{S. Stasevic\inst{1,2}, J. Milli\inst{2}, J. Mazoyer\inst{1}, A.-M. Lagrange\inst{1,2}, S. Bergeon\inst{2}}
\institute{
LIRA, Observatoire de Paris, Universit\'{e} PSL, Sorbonne, Universit\'{e}, Universit\'{e} Paris Cit\'{e}, CNRS, 5 place Jules Janssen, 92195 Meudon, France
\and
Univ. Grenoble Alpes, CNRS, IPAG, F-38000 Grenoble, France  
}

\date{Received 29 April 2025; accepted 20 July 2025}

%
%-------------------------------------------------------------------
% \abstract{}{}{}{}{} 
% 5 {} token are mandatory
 
\abstract
   {The direct detection of circumstellar discs through high-contrast imaging provides key insights into the history and dynamics of planetary systems. Pole-on discs, especially faint debris discs, are difficult to detect and require careful consideration during post-processing to remove stellar residuals from the data while preserving the disc signal. Reference-star differential imaging (RDI) serves as one of the primary post-processing methods for disc observations; however, the impact of the reference library on the detection sensitivity of discs has yet to be fully explored.}
   {We aim to explore different reference library selection metrics in order to develop a method of reference frame selection that is optimised for pole-on discs to be used for the upcoming large-scale RDI reduction of archival SPHERE/IRDIS observations in the search of new discs.}
   {We performed RDI post-processing based on principal component analysis on 20 targets without discs and with varying observational conditions and seven targets with discs, using reference libraries built from frames that were preselected to best match different observational, atmospheric, and stellar parameters of the science frames. The contrast of the disc-free reductions was measured, and forward modelling was used to estimate the signal loss from over-subtraction using synthetic pole-on discs with two different widths and four different radii. The signal-to-noise ratio (S/N) of the real disc targets was measured.}
   {Diverse reference libraries built using subsets of frames that closely matched different parameters achieved the best disc S/N and smallest deviation from the best contrast of each target, outperforming libraries built using a single criteria as a selection metric. Libraries built using frame-to-frame Pearson correlation coefficient alone as a selection criterion achieved the best mean contrast overall. Both selection metrics performed consistently well for all disc radii and observational conditions. We also found that reference libraries built using frames observed close in time to the science frame performed well for discs at small separations, giving the best contrast for $\sim$\,30\% of the targets at a radius of 20px.}
   {}


%\authorrunning{}
\titlerunning{Optimising reference library selection for disc RDI}
\maketitle
\nolinenumbers

%
%-------------------------------------------------------------------
\section{Introduction}
\label{sec:intro}

High-contrast imaging (HCI) is an observational technique that is paramount for the detection of wide separation giant planets and circumstellar discs. These observations further our understanding of disc dynamics \citep[e.g.][]{Lin2019, Kennedy2020}, planet-disc interactions \citep[e.g.][]{Espaillat2014, Dong2018, Pearce2024a}, and the formation and evolution of planetary systems \citep{Bowler2016, Follette2023}. Detections of circumstellar discs in particular offer key insights into the history and dynamics of a system in which they both influence and are influenced by planets that may be present there. While the analysis of spectral energy distributions (SEDs) -- the first detection method for discs -- provides information on the quantity and location of dust in a disc \citep{Pearce2024}, HCI observations help lift the degeneracy of SED modelling alone by directly probing their dust properties and morphology \citep{Esposito2020}. 

Debris discs are formed by second-generation dust: short-lived dust particles sustained through the continuous collision of larger parent bodies in a steady-state equilibrium \citep{Wyatt2002}. The optical depths of debris discs are much lower than those of primordial protoplanetary discs and are thus fainter and more difficult to detect. Within the population of resolved debris discs, morphologies have proven to be highly diverse \citep{Matra2025}, with many showing unique structures -- for example, warps, spirals, gaps, asymmetries, and clumps -- that could be attributed to interactions between planetary-mass companions and dust \citep{Lee2016, Hughes2018}. Directly imaging these features can help constrain the presence of companions that cannot yet be detected \citep{Terrill2023}. Improving detection capabilities increases our ability to resolve faint discs and structures and possibly detect the currently unseen companions residing amongst them.

One of the biggest challenges in HCI is the removal of stellar residuals that greatly hinder the achievable contrast of an image. Quasi-static speckles -- residual scattered starlight from small wavefront aberrations caused by imperfections in the imaging system \citep{Racine1999} -- can be difficult to distinguish from faint planets or planets embedded in protoplanetary discs. Large-scale structures, such as the stellar halo caused by adaptive optics (AO) residuals and those formed by starlight leakage from the central coronagraph \citep{Cantalloube2019}, are highly limiting for the detection of discs, as they can be mistaken for discs and obscure extended features. Differential imaging (DI) techniques allow us to separate the residual starlight from the planet and disc signal so it can be removed from the image.

Angular differential imaging \citep[ADI;][]{Marois2006} takes advantage of the field of view (FOV) rotation during an observing sequence of an altitude-azimuth telescope, in which the circumstellar signal rotates with the FOV, while the stellar noise does not. While using frames taken during the same observation of a target ensures that speckle noise is highly correlated, the object itself is included in the frames, which can lead to self-subtraction when performing ADI. Self-subtraction occurs when the astrophysical signal does not move a sufficient amount between subsequent frames and is thus included in the residual starlight model and subtracted from the image \citep{Esposito2014}. For planets this primarily affects the region close to the star, where the same angular rotation translates to a smaller change in position than at larger separations. Discs, however, face an additional challenge due to their geometry. As the inclination of a disc decreases, so does its rotational variance, leading to more disc flux being included in the model of stellar noise \citep{Milli2012}. This eventually leads to the case of a pole-on disc, which will be completely self-subtracted regardless of the separation from the star or how much parallactic rotation occurs through the observation sequence. Self-subtraction thus biases disc morphology and can even create artefacts in protoplanetary discs that can be mistaken for planets, as in the case of LkCa\,15 \citep{Currie2019}.

We therefore turn to reference-star differential imaging \citep[RDI;][]{SmithTerrile1984,Lecavelier1993,Kalas1995,Beuzit1997,Mouillet1997} for the reduction of pole-on discs in total intensity imaging. RDI, similar to ADI, uses observations to build a model of the speckle pattern to be subtracted. Instead of using frames from the science target observation sequence, RDI uses observations of other targets that do not contain any of the astrophysical signal we wish to preserve. By negating the effects of self-subtraction, RDI aids in the detection of new discs while also improving sensitivity at small separations \citep{Xie2022,Sanghi2024}.

While classically the scaled flux of a reference star is subtracted from the science target, algorithms such as principal component analysis \citep[PCA, also known as Karhunen-Lo\`eve transform;][]{Soummer2012,Amara2012} can be applied to better model the stellar noise and improve contrast. The aggressive subtraction of speckles, however, can lead to over-subtraction: the subtraction of astrophysical signal that has been mistaken for stellar noise by the algorithm \citep{Pueyo2016}. This not only affects the photometry of detected objects but also the shape of imaged discs, which could lead to the incorrect interpretation of extended features. Algorithms optimised for disc reductions, such as those employing non-negative matrix factorisation \citep[NMF;][]{Ren2018} and data imputation in Karhunen-Lo\`eve transforms \citep[DIKL;][]{Ren2023}, have been successful in reducing over-subtraction and thus better preserving disc morphology. Forward modelling techniques for discs have also been developed to reduce the effects of over-subtraction and, if applied to ADI reductions, self-subtraction \citep{Esposito2014,Pueyo2016}; however, these techniques require assumptions to be made on the disc morphology that may not be representative of the real disc. \cite{Mazoyer2020} have addressed this by repeating the forward modelling process many times while varying the initial disc model parameters, but this technique is still limited by the capabilities of the disc model in producing complex features.

In addition to benefiting from algorithms that improve the modelling of stellar noise, RDI reductions have seen an improvement through the utilisation of polarimetric observations to perform constrained-RDI \citep{Lawson2022}, the use of combined angular and reference-star differential imaging \citep[ARDI;][]{Juillard2024}, and the employment of the star-hopping method \citep{Wahhaj2021}. For the last two techniques, the key to their success in improving RDI reductions is the use of a reference library with frames that closely match the speckle noise of the science observation, with \citet{Juillard2024} showing that ARDI has the most marked improvement over RDI when the reference-star frames are of poor quality. Careful reference frame selection is therefore vital for optimising the reduction capabilities of RDI, especially for observations not performed in star-hopping mode, where the time between observations far exceeds the speckle lifetime. 

The use of image comparison metrics to select reference frames greatly improves the performance of RDI \citep{Ruane2019, Juillard2024, Romero2024}; however, there is no clear consensus on the best metric to use. For exoplanets, \citet{Ruane2019} find that the structural similarity index metric (SSIM) gives a slight improvement in signal-to-noise ratio (S/N) compared to the Pearson correlation coefficient (PCC). On the other hand, \citet{Romero2024} find the opposite, with reference frames selected based on the frame-to-frame PCC within the speckle-dominated region giving the best S/N compared to all other metrics tested. Meanwhile, \citet{Juillard2024} find that using the reference frames with the highest PCC in the ARDI library does not always outperform reductions using a library where these highly correlated frames are omitted. The use of a linear correlation, such as the PCC, which is greatly affected by pixel outliers, may not even be an appropriate metric for selecting reference frames if said outliers only appear in a small proportion of the reference library and thus have a minimal influence on the stellar noise model. Furthermore, different sources of wavefront aberrations produce unique noise structures \citep[see e.g.][]{Cantalloube2019}, which when combined for a single image, may not be well represented by any one reference frame. Building reference libraries using variables that either impact the structure of speckle noise or provide a measure of observational quality as selection metrics \citep{Romero2024} or with subsets of reference frames that match different noise substructures could thus improve speckle subtraction, with the added benefit of being independent of the algorithm chosen to model the noise.

We investigate the impact of reference library selection on a small sample of data taken with the Very Large Telescope/Spectro-Polarimetric High-contrast Exoplanet REsearch \citep[VLT/SPHERE;][]{Beuzit2019} in order to determine the optimum method to use for a large-scale reduction of SPHERE observations aimed at detecting new discs in archival data. We evaluate the performance of different libraries built by matching observational properties, stellar properties, or atmospheric conditions on the RDI reduction of a set of 20 disc-free targets using synthetic discs and seven targets with previously resolved discs. For the synthetic disc analysis, we used models of pole-on debris discs. Debris discs are fainter and more difficult to detect than protoplanetary discs, and pole-on discs are completely self-subtracted when reduced with ADI. As such, these are the targets that benefit the most from improved RDI performance. 

In Sect.~\ref{sec:data} we describe the observations and various metadata used in the study and detail the science targets and methods for reference frame selection, RDI reduction, and performance assessment in Sect.~\ref{sec:method}. We present the results of the systematic analysis of disc-free targets using synthetic disc forward modelling in Sect.~\ref{sec:results} and the performance on reductions of known discs in Sect.~\ref{sec:known_discs} before concluding in Sect.~\ref{sec:concl}.

\section{Data}
\label{sec:data}

\subsection{Observations}
\label{sec:d_obs}

We used archival total intensity observations taken with the VLT/SPHERE using the Infrared Dual-band Imager and Spectrograph \citep[IRDIS;][]{Dohlen2008,Vigan2010} in pupil-stabilised mode. Astrometric calibrations of these data were performed on-sky \citep[as detailed in][]{Maire2016}, using a true north correction and updated IRDIS plate scales from \cite{Maire2021}. The images cover an 11\arcsec{}\,$\times$\,11\arcsec{} FOV.

Our available dataset consisted of 1298 observation sequences taken with the DB\_H23 filter (centred at 1.586\micron{} and 1.666\micron{} with filter bandwidths of 0.053\micron{} and 0.056\micron{} respectively) between 2015-02-03 and 2023-07-15, and 884 observation sequences taken with the DB\_K12 filter (centred at 2.103\micron{} and 2.254\micron{} with filter bandwidths of 0.101\micron{} and 0.110\micron{} respectively) between 2014-07-16 and 2023-07-25. All observations were taken with the star placed behind the N\_ALC\_YJH\_S apodised Lyot coronagraph and a 185\,mas wide focal mask. 

The raw data were processed by the High-Contrast Data Centre \citep[HC-DC\footnote{\href{https://hc-dc.cnrs.fr}{https://hc-dc.cnrs.fr}};][]{Delorme2017}, which implements the SPHERE Data Reduction and Handling pipeline \citep[DRH;][]{Pavlov2008}. It performs dark, flat, and bad pixel corrections of the frames and registers the coronagraphic images to a common centre. We note that in this paper, we use the term `frame' to refer to a single exposure image and `cube' to refer to a full observing sequence.

\subsection{Metadata}
\label{sec:d_meta}

Data other than the IRDIS images, coming from atmospheric, meteorological, and telescope sensors, were used in this work. We refer to these data, and the stellar parameters of the targets, as `metadata'.

Many factors affect the structure of noise in ground-based HCI observations, with the dominant factor limiting contrast varying with separation \citep[e.g.][]{Kasper2012}. Outside of the AO correction region, the noise is dominated by the atmospheric seeing. Within the AO correction region, quasi-static speckles arising from non-common path aberrations can become a dominant source of noise under good observing conditions, their intensity increasing with longer exposure times and brighter targets \citep{Lafreniere2007}. Target brightness also contributes to the strength of low-order atmospheric residuals, as wavefront sensing is noisier for fainter targets, leading to poorer correction \citep{Cantalloube2019}. While discs are easy to distinguish from point-like speckles prior to the application of post-processing techniques, self- and over-subtraction can cause discs to appear patchy, and unsubtracted speckles could be mistaken for disc signal in the reduced image. 

In addition to atmospheric and stellar factors, weather conditions can affect the shape of the stellar point-spread function (PSF), leading to unique noise structures. High wind speed turbulent layers give rise to a wind driven halo (WDH, also known as AO servolag error), in which the PSF is elongated along the direction of the wind at the telescope pupil, creating a butterfly pattern in the coronagraphic image \citep{Cantalloube2018,Madurowicz2019,Cantalloube2020}. This effect is strongly correlated with the high-altitude jet stream layer at $\sim$\,200mbar (12km above sea level) and exhibits a brightness asymmetry due to interference between servolag and scintillation effects \citep{Cantalloube2018}. On the other end of the scale, low wind speeds in the telescope dome can also modify the shape of the PSF. Dubbed the low-wind effect \citep[LWE;][]{Sauvage2015}, these conditions are badly handled by the AO system \citep{Pourre2022}, forming secondary lobes around the core of the PSF that are very detrimental to the contrast achieved in coronagraphic imaging \citep{Milli2018}. The LWE is present in SPHERE observations taken before 2017-08-08 when the wind speed at 30m drops below 3m/s. After application of a coating to the telescope spiders \citep[described in][]{Milli2018}, the LWE only occurs for 30m wind speeds below 1m/s. Disc detection is highly limited by the large structures created by these AO residuals, which can both obscure and be mistaken for extended disc features. 

Target elevation plays a role in both the performance of the AO system and the stability of quasi-static speckles. At higher elevations, atmospheric seeing becomes lower, and thus AO residuals are expected to decrease \citep[see Sect. I.2.4 of][]{Roddier1999}. Quasi-static speckles are also most stable at small hour angles (i.e. high elevation), as both the altitude tracking speed and derotator speed are minimised when the target is highest in the sky, thus minimising optical aberrations induced by moving optical components and mechanical flexures of the telescope \citep[see Appendix D of][]{Milli2016}.

We selected ten metadata parameters that impact noise and observation quality to use as metrics for selecting reference frames. The metadata and their sources are listed in Table~\ref{tab:params}. The 200mbar wind direction was corrected by the parallactic angle of each frame to give the on-frame wind direction, and spectral type (SpT) was converted to a numerical value (i.e. O1\,$\equiv$\,11, M9\,$\equiv$\,79). Elevation was computed using the \texttt{AltAz} module of the Astropy Python package \citep{Astropy2022} to convert right ascension and declination to horizontal coordinates at the time of observation.
The stellar metadata were obtained for each unique target, and the other metadata for each observation. For the parameters whose measurements are independent of the HCI observation, the timestamp file -- output by the DRH data processing -- was used to obtain the metadata for each individual frame by matching the frame observation time to the closest parameter sampling time. For the DIT and stellar parameters, a single value was used for the full cube. 

\begin{table}[ht]
    \centering
    \begin{threeparttable}
        \caption{Metadata used as RDI reference library selection metrics.}
        \label{tab:params}
        \begin{tabular}{l|l|l}
            \hline
            Parameter & Source & Ref. \\
            \hline
            Seeing (\arcsec{})& Paranal ASM database$^{(\text{a,c})}$ & (1,2,3) \\
            Coherence time, $\tau_0$ (s)& Paranal ASM database$^{(\text{b,c})}$ & (1,2,3) \\
            30m wind speed (m/s)& Paranal ASM database$^{(\text{d})}$ & (1) \\
            200mbar wind (m/s; $^\circ$)& ECMWF CDS catalogue$^{(\text{e})}$ & (4,5) \\
            Elevation ($^\circ$) & Recomputed DRH output & \\
            Epoch (mjd) & DRH output &  \\
            DIT (s) & FITS header &  \\
            G magnitude, M$_\text{G}$ & Gaia DR3 catalogue & (6) \\
            H magnitude, M$_\text{H}$ & 2MASS All Sky catalogue & (7) \\
            Spectral type, SpT & SIMBAD database & (8) \\
            \hline
        \end{tabular}
        \tablefoot{
            The sources of the data are noted for each parameter. The seeing is measured by the Differential Image Motion Monitor (DIMM) at 0.5\micron{}. Coherence time after 2016-04-05 has additional measurements from the Multi-Aperture Scintillation Sensor (MASS). The DIT (detector integration time) refers to a single exposure time.\\
            \tablefoottext{a}{Astronomical site monitor (ASM) DIMM seeing sampled every 79s.}\\
            \tablefoottext{b}{MASS-DIMM combined measurements sampled every 79s.}\\
            \tablefoottext{c}{Measurements before 2016-04-05 accessed from historical ambient data; sampled every 60s with DIMM3.}\\
            \tablefoottext{d}{Vaisala meteorological station measurements sampled every 60s.}\\
            \tablefoottext{e}{European Centre for Medium-Range Weather Forecasts (ECMWF) Climate Data Store (CDS). ECMWF re-analysis (ERA5) sampled every hour.}
        }
        \tablebib{
            (1)~\url{https://archive.eso.org/wdb/help/eso/ambient_paranal.html}; (2)~\cite{Kornilov2007}; (3)~\cite{Sarazin1990}; (4)~\cite{C3S2023}; (5)~\cite{Hersbach2020}; (6)~\cite{GaiaCollaboration2020}; (7)~\cite{Cutri2003}; (8)~\cite{Wenger2000}
        }
    \end{threeparttable}
\end{table}

\section{Methods}
\label{sec:method}

\subsection{Sample selection}
\label{sec:m_science}

For the systematic contrast analysis, we selected a set of targets from the H23 dataset described in Sect.~\ref{sec:d_obs} without any known discs resolved with HCI, having either no infrared excess reported by \cite{McDonald2017} or, if flagged as a candidate for hosting infrared excess, having a fractional luminosity below $10^{-5}$, as this is below the current disc detection limit for HCI scattered light observations. Targets were grouped by observing condition, having either LWE, WDH, or good, average, or bad seeing with no wind effects. A total of 20 observation sequences were selected, each with a DIT (detector integration time) of 32s, distributed evenly between early-type (A1--B8) and late-type (F4--K2) stars and the different observing conditions. 

In addition to the sample above, seven targets with known discs were selected for testing, all having previously been resolved in scattered light with SPHERE. The discs of HD\,111520 and HR\,4796 are edge-on and of intermediate inclination, respectively, with the remaining discs being pole-on, or close to pole-on. Both protoplanetary and debris discs were included in the sample. 

The selected targets and observation sequences (referred to as `science' targets or cubes) are summarised in Table~\ref{tab:obs_sci} alongside the reason for their selection. The metadata of each science cube is presented in Table~\ref{tab:apx_obs_sci_full}.

\begin{table}[ht]
    \centering
    \begin{threeparttable}
        \caption{Observation sequences used as science targets in this study.}
        \label{tab:obs_sci}
        \begin{tabular}{l|l|l|l}
            \hline
            Target name & Observation night & Condition & Disc ?\\
                & YYYY-MM-DD & & \\
            \hline
            2MASS J1604$^{(\text{a})}$ & 2015-06-09$^{(\text{b})}$ & Bad seeing & Y\,(1) \\
            BD-19 4288 & 2015-04-28 & LWE & N \\
            HD 9054 & 2016-11-16 & Bad seeing & N \\
            HD 13724 & 2018-08-18 & Good seeing & N \\
            HD 16978 & 2016-09-15 & Good seeing & N \\
            HD 37306 & 2015-10-28 & Average & N \\
            HD 37484 & 2017-12-01 & Bad seeing & N \\
            HD 59967 & 2018-03-27 & Average & N \\
            HD 69830 & 2018-12-18 & Average & N \\
            HD 100453 & 2016-01-19 & Bad seeing & Y\,(2) \\
            HD 107301 & 2016-01-18 & Bad seeing & N \\
            HD 111520 & 2016-05-14 & WDH & Y\,(3) \\
            HD 113902 & 2016-06-05 & WDH & N \\
            HD 121156 & 2017-02-05 & Good seeing & N \\
            HD 123247 & 2015-04-05 & LWE & N \\
            HD 126062 & 2016-04-08 & Average & N \\
                    & 2016-07-23 & WDH & N \\
            HD 126135 & 2018-03-16 & Good seeing & N \\
            HD 128987 & 2016-05-31 & WDH & N \\
            HD 141569 & 2015-05-15 & Average & Y\,(4) \\
            HD 147808 & 2016-04-14 & LWE & N \\
            HD 165185 & 2019-09-06 & WDH & N \\
            HD 181327 & 2015-05-09 & Bad seeing & Y\,(5) \\
            HD 188228 & 2015-06-30 & LWE & N \\
            HD 213398 & 2016-06-13 & Bad seeing & N \\
            HR 4796 & 2015-02-02 & Average & Y\,(6) \\
            TW HYA & 2015-02-03 & Average & Y\,(7) \\
            \hline
        \end{tabular}
        \tablefoot{
            The observing condition and presence of a resolved disc are noted for each science cube. References are given for the first scattered light observations of the disc targets. Note that the observing conditions of the disc targets are listed here for completion, but are not taken into account in the analysis.\\
            \tablefoottext{a}{2MASS J16042165-2130284.}\\
            \tablefoottext{b}{Observation taken with DB\_K12 filter.}
        }
        \tablebib{
            (1)~\cite{Mayama2012}; (2)~\cite{Wagner2015}; (3)~\cite{Padgett2016}; (4)~\cite{Augereau1999, Weinberger1999}; (5)~\cite{Schneider2006}; (6)~\cite{Schneider1999}; (7)~\cite{Krist2000}
        }
    \end{threeparttable}
\end{table}

\subsection{RDI reduction}
\label{sec:m_rdi}

From the dataset described in Sect.~\ref{sec:d_obs}, we removed the observing sequences of targets with known discs, planets, or binaries and selected all observations taken without satellite spots to populate our library of potential reference frames \citep[hereafter called the `master reference library', following the nomenclature used in][]{Xie2022}. The H23 master reference library contained 67\,386 frames from 769 observation sequences, and the K12 master reference library contained 54\,773 frames from 746 observation sequences. For the disc-free science targets, observations of the target taken at any epoch were removed from the master reference library when building the final libraries for that target. The PCC was computed between the mean subtracted frames of the science cubes and master reference library for the same wavelength channel. The calculation was performed within a circular annulus between 0.18\arcsec{} and 0.43\arcsec{}; where the speckle noise dominates.

For this study, we chose to build unique reference libraries for each science frame and wavelength channel, consisting of 1000 reference frames each. This value was selected in order to compromise between the computational storage required for the libraries and the dependence between contrast and reference library size, which increases for an increasing number of reference frames \citep{Ruane2019,Xie2022}. For all parameters listed in Table~\ref{tab:params} except DIT, the absolute difference between the science frame value and reference frame values was calculated. The 10\,000 reference frames with parameter values most closely matching that of the science frame were preselected, equating to $\sim$\,15\% and $\sim$\,18\% of the H23 and K12 master reference libraries, respectively. This ensured that the parameter values of the reference libraries would be similar enough to the science that the choice of parameter was the variable being tested, while also maintaining a large enough frame sample that the final reference libraries still contained well-correlated frames. Additionally, only reference targets of the same luminosity class as the science target were considered for the SpT library. For the DIT library, all reference frames sharing the same DIT as the science target were preselected. The final reference library for each science frame was then built using 1000 preselected reference frames with the highest PCC. 

In addition to the ten individual parameter libraries, mixed libraries were built, taking 100 reference frames with the highest PCC from 5000 preselected frames for each of the ten parameters. For comparison purposes, reference libraries were also built with no preselection -- using only PCC as a selection metric -- and random selection from the master reference library. This gave a total of 13 reference libraries tested for each science target.  

After building the reference libraries, each science target was reduced using the principal component analysis (PCA) technique described in \cite{Soummer2012} and \cite{Amara2012}. The science and reference frames were cropped to a 3.1\arcsec{}\,$\times$\,3.1\arcsec{} (256\,$\times$\,256px) field of view and the area covered by the coronagraph was masked using a central, 0.1\arcsec{} (8px) radius aperture. Each frame was then subtracted by its mean value. The disc-free targets were reduced with 1--500 principal components (PCs) in steps of 20 (barring the first step of 19), and the disc targets with 50--300\,PCs in steps of 50. After PCA subtraction, the frames were derotated by their parallactic angles and median-combined along the temporal and spectral axes to form a single reduced image for each PC value used.

\subsection{Throughput corrected contrast}
\label{sec:m_contrast}

The contrast of the disc-free science cube reductions was assessed at separations of 0.25\arcsec{}, 0.49\arcsec{}, 0.74\arcsec{}, and 0.98\arcsec{} (20px, 40px, 60px, and 80px). For the H23 wavelength filter, 0.25\arcsec{} and 0.49\arcsec{} correspond approximately to the inner and outer radii of the speckle-dominated region, and 0.74\arcsec{} and 0.98\arcsec{} correspond to radii just within and outside of the AO correction region. The contrast was measured using a set of non-overlapping apertures of diameter $\frac{\lambda}{D}$ (where $\lambda$ is the wavelength of the observation and $D$ is the diameter of the telescope; equal to 0.04\arcsec{} (3.35px) for the H23 filter), centred along the circumference of a circle with radius equal to the given separation. The sum of the flux in each aperture was measured, and the standard deviation calculated. The standard deviation was multiplied by 5, and divided by the sum of the central stellar PSF within an aperture of 0.04\arcsec{} diameter to give the uncorrected 5-sigma contrast.

We used forward modelling to estimate the signal loss due to over-subtraction that would occur for a pole-on disc. In this paper, we refer to this signal loss with the term `throughput', which therefore refers to a post-processing throughput, and not the instrumental throughput of the signal. Synthetic scattered light discs were created using the \texttt{scattered\_light\_disk} and \texttt{fakedisk} modules of the Vortex Image Processing (VIP)\footnote{\url{https://github.com/vortex-exoplanet/VIP}.} Python package \citep{Christiaens2023, Gonzalez2017, Augereau1999} with a maximum flux of 1~count/s, inclination of 0$^{\circ}$, and semi-major axes of 20px, 40px, 60px, and 80px. For each semi-major axis, two discs were generated: a narrow disc with a width of 0.07\arcsec{} (6px; equal to a disc that is just resolved) and a wide disc with a width equal to half its semi-major axis. The discs were convolved with the stellar PSF observed at the same time as the science cube being assessed. Each synthetic disc image was projected on the PCs of a given reference library, and the projection was subtracted from the original disc to reproduce over-subtraction. We then divided this image by the original disc model to calculate the throughput.

The median of the throughput within an annulus with a width and radius equal to the disc was calculated. Since each science frame and wavelength channel used a different set of reference frames, the final throughput of a reduction was taken as the median of the individual science frame and channel throughput measurements. The contrast was divided by the throughput, and the PC with the best throughput-corrected contrast (henceforth simply referred to as `contrast') was selected for the final value. 

\subsection{Disc signal-to-noise ratio}
\label{sec:m_snr}

For the targets with known discs, the S/N of the disc after RDI reduction was used to quantify the performance of the reference libraries. To compute the noise of the reduced images, the region containing the disc was masked for each target. The standard deviation of the non-masked pixels was measured within concentric annuli, then multiplied by 5 to give the 5-sigma noise as a function of separation. The width of each annulus was set to be equal to the integer value of the full width at half maximum (FWHM) of the stellar PSF (4px for H23 data and 6px for K12 data). To account for the small sample statistics, a penalty term was calculated at the radius of each annulus following the strategy described in Sect.~3.4 of \citet{Mawet2014}, using a 5-sigma threshold as the confidence level and the FWHM as the size of a resolution element. A 1-dimensional linear profile was fit to the penalty-corrected noise, from which a 2-dimensional azimuthally averaged noise map was created.

The reduced images were divided by their corresponding noise maps to create S/N maps, and the mean S/N was  measured within the region of brightest disc flux. The same region was used for each PC and reference library reduction of a science target and so was defined, via visual inspection, to encompass the brightest disc regions of each reduced image while limiting the inclusion of negative signal caused by over-subtraction. The regions are shown in Appendix~\ref{apx:disc_reg}, overlaid on the RDI reductions of each disc using the different reference libraries. The PC with the best S/N for each selection metric was selected for the final value, with $\sim$\,50\% of all libraries having the best S/N when reduced with 50--100\,PCs.

\section{Results of systematic analysis}
\label{sec:results}

\subsection{Main findings}
\label{sec:r_main}

The reference library selection metric used for conducting the large-scale reduction must be independent of disc radius and width, as these are parameters we cannot know for potential systems that have not yet been resolved in scattered light. Thus our focus is on the performance of the different selection metrics for the science targets overall and for the different observing conditions, which we know prior to reduction.

\begin{figure}
    \centering
    \includegraphics[width=8.5cm]{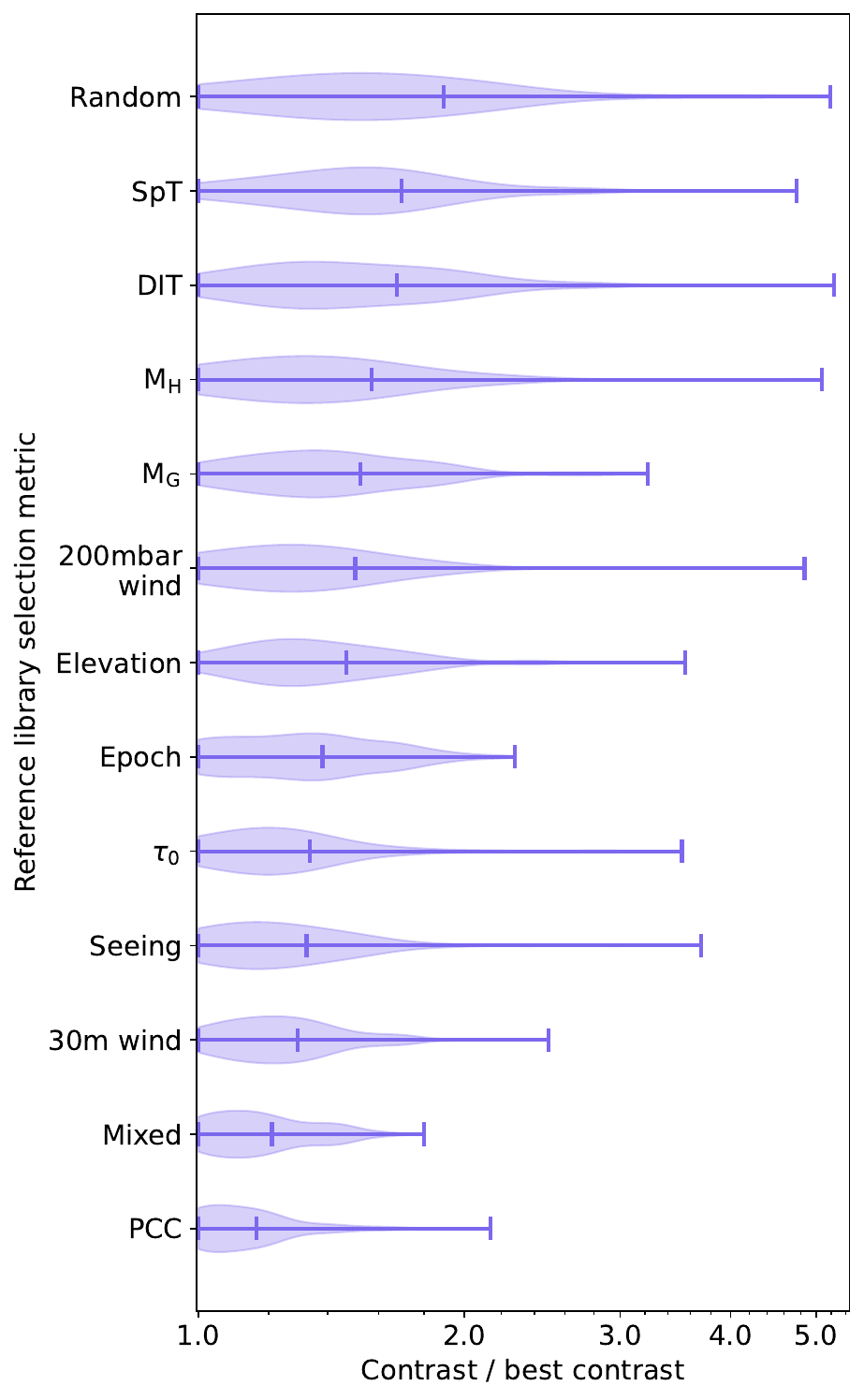}
    \caption{Violin plot showing the contrast distribution of each reference library selection metric for all science targets, disc radii, and widths. The contrast was normalised by that of the parameter library giving the best performance under the same measurement conditions. The vertical lines of the violin show the minimum, mean, and maximum values. Reference library selection metrics are ordered from top to bottom by descending mean.}
    \label{fig:viol_contrast}
\end{figure}

For each selection metric, the contrast was computed for the RDI reductions of the 20 disc-free cubes, using two different disc widths with four different radii, as explained in Sect.~\ref{sec:m_contrast}. The contrast measurements were normalised by the contrast of the reference library giving the best performance for the same target, disc width, and radius, yielding a distribution of normalised contrast values for each selection metric, described by the shaded curve in Fig.~\ref{fig:viol_contrast}. It should be noted that `better contrast' refers to a smaller contrast value and, as such, a smaller normalised contrast, with the best contrast normalised to 1.

Reductions using reference libraries built with only PCC as a selection metric yield the best contrast on average, with a mean normalised contrast of 1.16 and a maximum of 2.14. The mixed parameter libraries give the second best mean normalised contrast of 1.21, with a maximum of 1.80. While the PCC library achieves a lower mean than the mixed library, the difference between the maximum normalised contrast values shows that the mixed reference library has a more consistently high performance than the PCC library and is, in fact, the only selection metric to always achieve a contrast within a factor of 2 from the best contrast for any given target, separation, or disc width. In addition, building a mixed library requires significantly less computation time than building a PCC library. By preselecting frames from the master reference library before carrying out the frame-to-frame PCC calculations, we not only decrease the number of computations that need to be performed, but we also reduce the number of data cubes that need to be assessed and that have a finite read-in time.

Libraries built from randomly selected reference frames give the poorest overall performance, with a mean normalised contrast of 1.89 and a maximum of 5.18. Using reference library selection metrics thus improves RDI contrast performance. That being said, the SpT and DIT libraries only yield marginally better contrast than the random library on average, which suggests that they are not useful choices of selection metric, as they perform as well as if no selection metric was used at all.

\subsection{Target observation condition}
\label{sec:r_obs_type}

\begin{figure*}
    \centering
    \includegraphics[width=16cm]{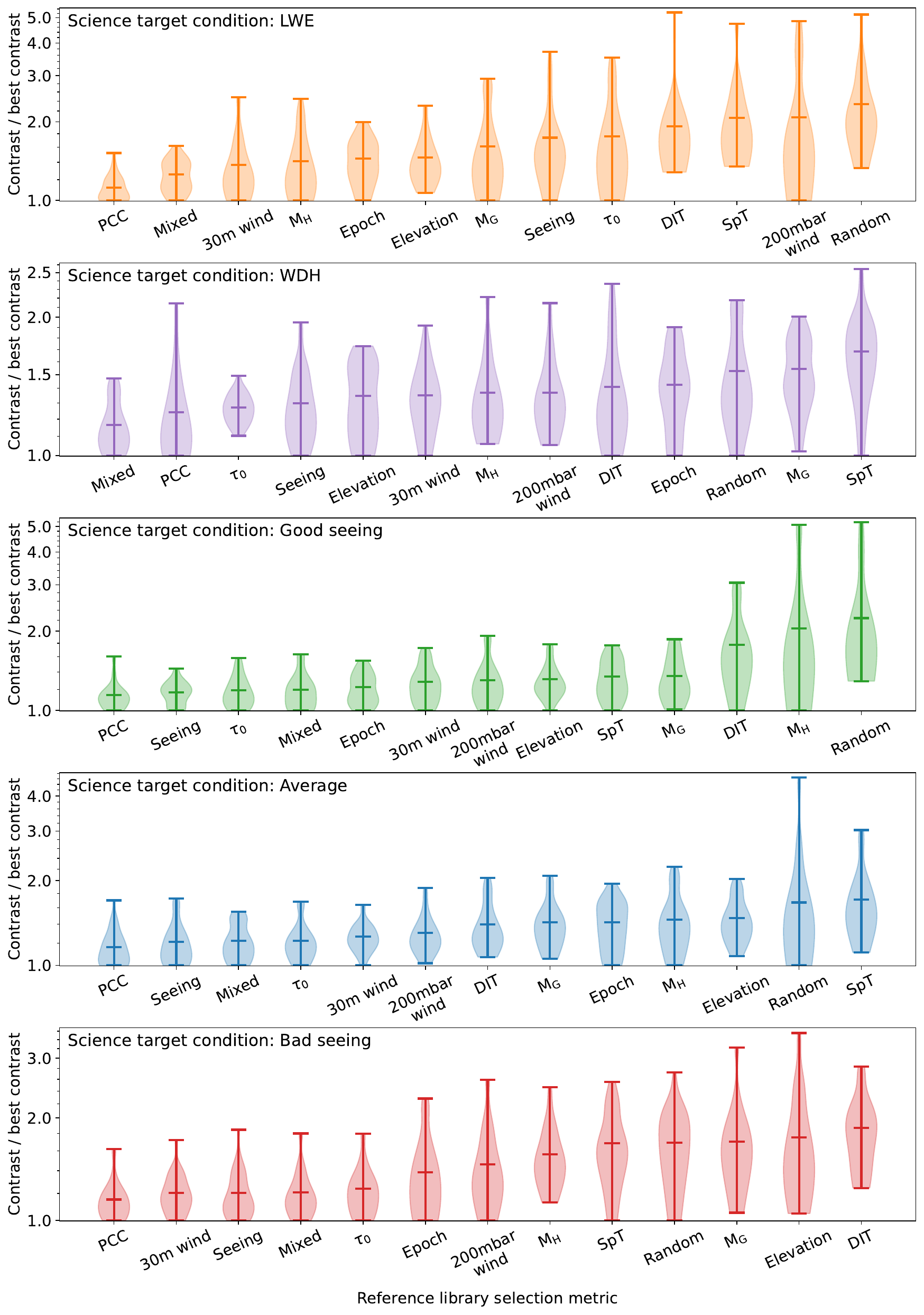}
    \caption{Violin plots showing the contrast distribution of each reference library selection metric for different observing conditions of the science targets. Plots show the distribution for the LWE (orange; top), WDH (purple), good seeing (green), average seeing (blue), and bad seeing (red; bottom) science targets, including all disc radii and widths. The contrast was normalised by that of the parameter library giving the best performance under the same measurement conditions. The horizontal lines of the violin show the minimum, mean, and maximum values. For each plot, reference library selection metrics are ordered from left to right by ascending mean.}
    \label{fig:viol_obs_contrast}
\end{figure*}

The normalised contrast distributions for each science target observing condition are shown in Fig.~\ref{fig:viol_obs_contrast}. We see that the choice of reference library selection metric has the largest impact on reductions of observations with good seeing or LWE, improving contrast up to a factor of $\sim$\,5. Science targets with WDH are impacted the least, with up to a factor of $\sim$\,2.5 improvement in contrast. This is to be expected, as images obtained under good observing conditions are mainly limited by speckle noise, and so the contrast is much more dependent on how well the reference library is able to describe the speckles so they can be subtracted through RDI. 

The mixed and PCC libraries give the best mean normalised contrast for science targets suffering from LWE and WDH. For the targets without wind effects, the PCC library gives the best mean normalised contrast, followed by the seeing library for targets with good to average seeing, and the 30m wind library for targets with bad seeing. The significance of the library rankings, however, must be examined. For the targets with good, average, and poor seeing, the mean normalised contrasts of the five best-performing selection metrics fall within 10\% of each other. Furthermore, the PCC, seeing, and mixed libraries yield the best contrast within $\sim$\,5\% for all three of the non-wind effect groups. As such, a meaningful conclusion on the best library to use for observations that do not suffer from wind effects cannot be drawn from this contrast analysis alone. 

For the LWE science targets, using the 30m wind velocity as a selection metric yields the best contrast after the PCC and mixed libraries, which we expect given the relationship between the two, discussed in Sect.~\ref{sec:d_meta}. Surprisingly, the M$_\mathrm{H}$ library also performs better, relative to the other libraries, for targets with LWE than for the other observing condition groups. The mean normalised contrast of the M$_\mathrm{H}$ library, however, does not vary significantly between the different observing condition groups, being in the range of 1.37--1.56 (1.41 for the LWE group), with the exception of the science targets with good seeing. This suggests that, rather than the M$_\mathrm{H}$ library performing better for LWE targets, the other selection metric libraries simply perform worse. 

It is also interesting to note that two of the selection metrics that perform worse for the LWE targets compared to the other targets are seeing and $\tau_0$. LWE occurs more frequently in good seeing and long $\tau_0$ conditions \citep{Milli2018}, with the latter being directly related to wind velocity through the equation $\tau_0 = 0.314 r_0/v_\text{eff}$, where $r_0$ is the Fried diameter and $v_\text{eff}$ is the effective wind velocity \citep{Roddier1981}. Observations suffering from LWE may therefore be indistinguishable from those with no wind effects and excellent observing conditions when only considering seeing and $\tau_0$. Thus, by selecting reference frames that closely match these two variables, we may over-represent high-quality observations that do not necessarily match the LWE noise structure well, especially at small separations from the star, yielding a poorer noise subtraction and hence, contrast.

The WDH science targets, on the other hand, do not achieve significantly better contrast when reduced with the 200mbar wind library compared to the other selection metrics, contrary to what we would expect. However, we can see that the $\tau_0$ library achieves the best mean normalised contrast after the mixed and PCC libraries, as well as one of the smallest maximums, always yielding a contrast within a factor of 1.5 from the best contrast. While the 200mbar wind library uses both the speed and on-frame direction of the wind when preselecting well-matching reference frames, $\tau_0$, and thus the $\tau_0$ library, only encompasses the speed. This suggests that wind direction may be a less important factor for determining well-matching reference frames for observations with WDH. Alternatively this could also mean that the 200mbar wind speed and direction provided by the ECMWF reanalyses are not always accurate, or that the WDH does not always correlate with that wind layer. In addition, when preselecting frames for the WDH libraries, we did not account for the scintillation, which governs the asymmetry of the WDH pattern. Since the strength of the asymmetry increases for stronger scintillation (and smaller wind speed), the WDH of two observations with the same 200mbar wind speed and direction may not be well matched if the scintillation is different, thus yielding a poorer noise subtraction. 

\subsection{Disc width and radius}
\label{sec:r_disc_size}

\begin{figure}
    \centering
    \includegraphics[width=8.5cm]{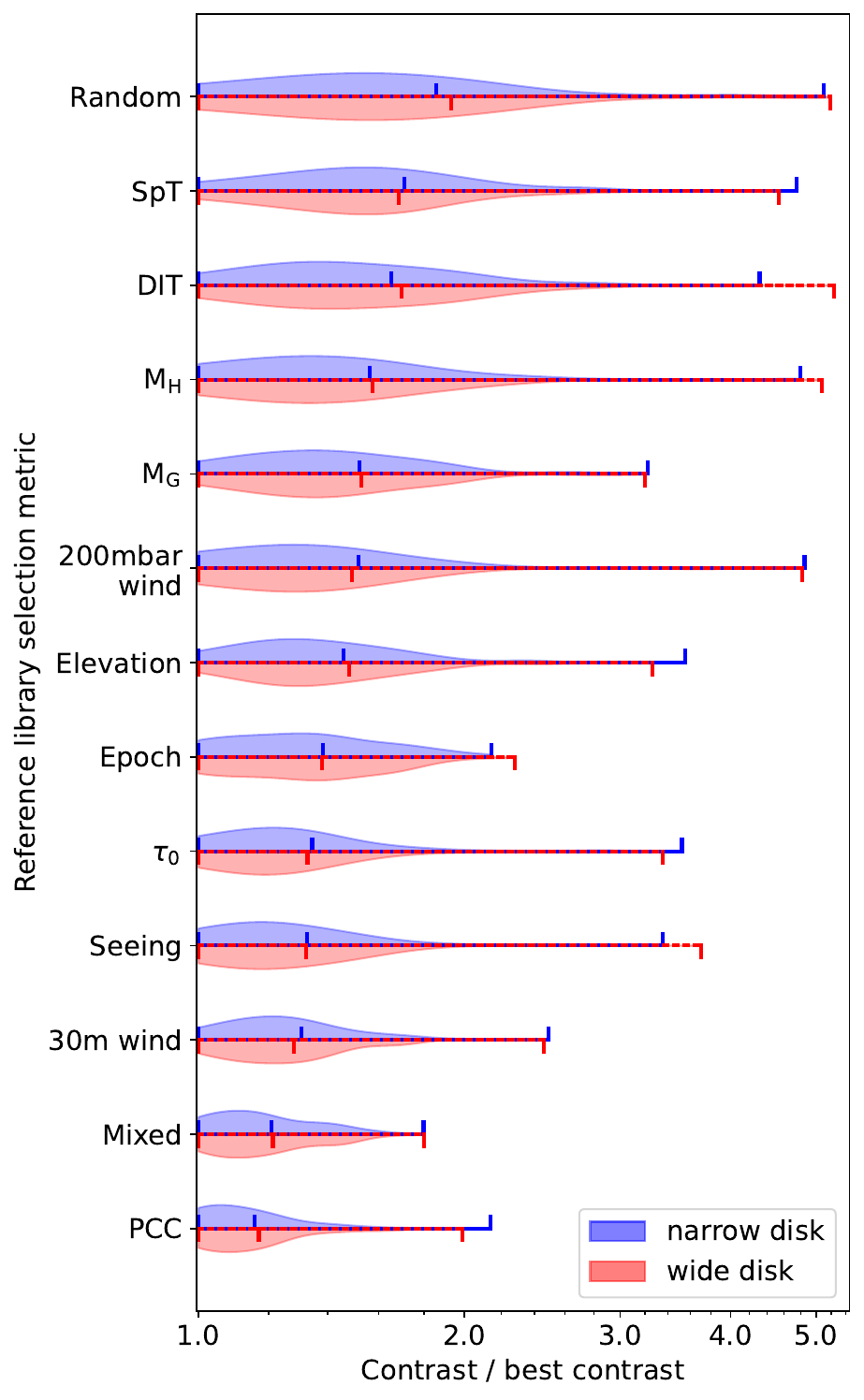}
    \caption{Violin plot showing the contrast distribution when throughput is calculated using a narrow (blue, upper side) and wide (red, lower side) synthetic disc for all science targets and disc radii. The contrast was normalised by that of the parameter library giving the best performance under the same measurement conditions. The vertical lines of the violin show the minimum, mean, and maximum values. Reference library selection metrics are ordered from top to bottom by descending mean.}
    \label{fig:viol_width_contrast}
\end{figure}

While the study was performed with the aim of finding a reference library selection metric that performed well regardless of disc size, assuming that for the large-scale reduction of SPHERE/IRDIS data these would be unknown, the relationship with disc radius and width is useful for improving RDI detection sensitivity for known discs. 
Figure~\ref{fig:viol_width_contrast} shows the normalised contrast distributions of the narrow and wide discs. We can see that, for each selection metric, the distribution is very similar between the two disc widths. The optimal reference library selection metric therefore remains the same regardless of the width of the disc we wish to reduce. It should be noted that the throughput, and thus the unnormalised contrast, does vary with disc width, but the factor by which it does so remains approximately constant for the different reference libraries. The narrow discs yield a higher throughput than the wide discs by a mean factor of 1.42\,$\pm$\,0.13, 1.91\,$\pm$\,0.19, 1.52\,$\pm$\,0.09, and 1.11\,$\pm$\,0.06 for the 20px, 40px, 60px, and 80px radius discs, respectively, with the uncertainties given by the standard deviation across the parameter libraries.

\begin{figure*}
    \centering
    \includegraphics[width=16cm]{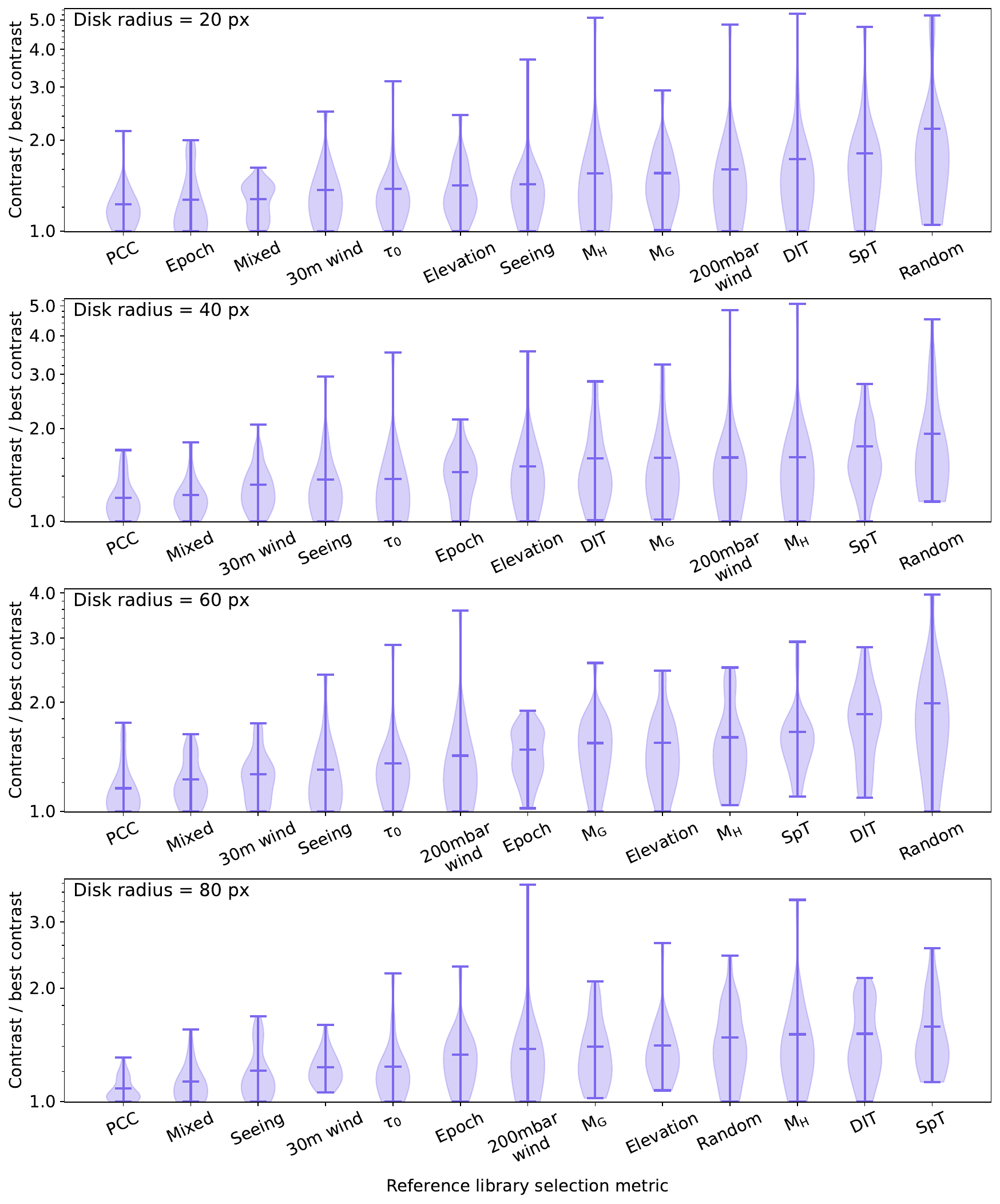}
    \caption{Violin plots showing the contrast distributions measured at a 20px (top), 40px, 60px, and 80px (bottom) separation for all science targets and disc widths. The contrast was normalised by that of the parameter library giving the best performance under the same measurement conditions. The horizontal lines of the violin show the minimum, mean, and maximum values. For each plot, reference library selection metrics are ordered from left to right by ascending mean.}
    \label{fig:viol_radius_contrast}
\end{figure*}

The contrast distribution for each disc radius is shown in Fig.~\ref{fig:viol_radius_contrast}. For radii above 20px, the relative performance of the different selection metrics largely follows that of the overall distribution, with the PCC and mixed libraries giving the best mean normalised contrast. For the 20px radius discs, the epoch library achieves the second best mean normalised contrast, differing from that of the PCC library by only 4\%. It also yields the best contrast for 13 of the 40 configurations of science target and disc width at this separation, while the next highest frequency is five configurations for the PCC library. Speckles are likely the dominant source of noise in this region, as speckle density increases at smaller separations, and so are more well matched for observations taken closely in time. Additionally, the centring of stars behind the coronagraphic mask is more consistent for observations taken in the same night, and thus the stellar residuals, which are affected by shifts in the position of the mask, are also more consistent.
We also see that the performance of the seeing library improves with increasing disc radius. While seeing only performs better than half the selection metrics for the 20px discs, it achieves the third best contrast for the 80px radius discs. This is expected given that noise becomes seeing dominated outside of the AO correction radius, which for H23 observations, is at a separation of $\sim$\,65px.

\section{Performance on real discs}
\label{sec:known_discs}

\begin{figure}
    \centering
    \includegraphics[width=8.5cm]{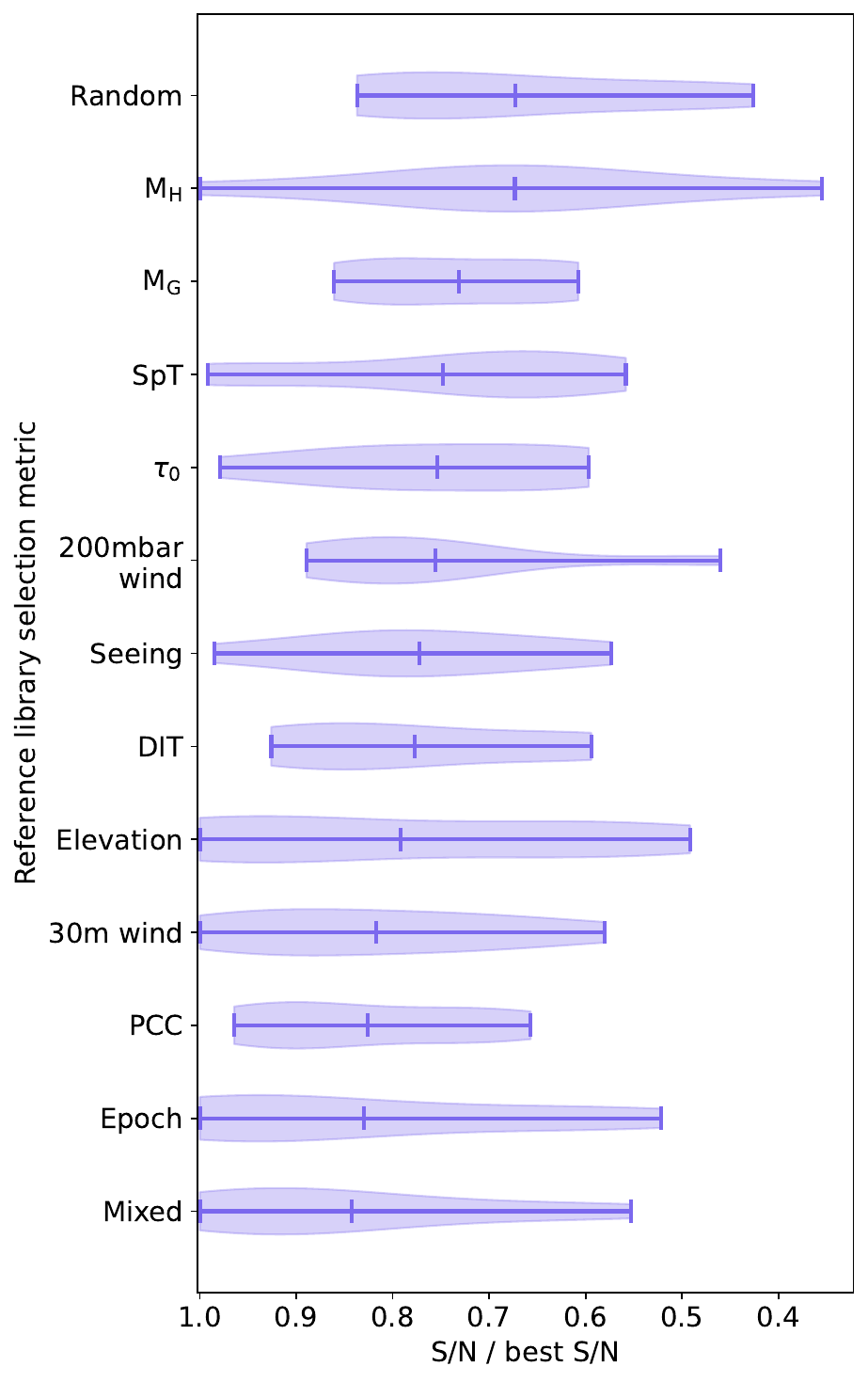}
    \caption{Violin plot showing the distribution of disc S/N measured on RDI reductions using reference libraries built with different selection metrics. The S/N measurements were normalised by the best S/N achieved for that disc using any reference library. The reference library selection metrics are ordered from top to bottom by ascending mean.}
    \label{fig:viol_disc}
\end{figure}

The S/N of the discs was measured for each reference library reduction as described in Sect.~\ref{sec:m_snr} and normalised by the best S/N achieved for that disc. Fig.~\ref{fig:viol_disc} shows the normalised S/N distribution of each reference library selection metric for all seven resolved disc targets.

The mixed library performs the best overall, with a mean normalised S/N of 0.84 and a minimum of 0.55. Reductions using the epoch and PCC libraries both give the second highest mean normalised S/N of 0.83, and minimum ratios of 0.52 and 0.66, respectively. Only the epoch and elevation libraries yield the highest S/N for more than one target. The epoch library performs much better, relative to the other selection metrics, for disc S/N than it does in the overall contrast distribution shown in Sect.~\ref{sec:r_main}. This is unsurprising, as the disc sample is biased towards discs with small angular radii, with the mean separation of the S/N measurement region being <40px (0.49\arcsec{}) for five out of seven targets (which can be seen in the figures in Appendix~\ref{apx:disc_reg}). It is therefore in agreement with the contrast results discussed in Sect.~\ref{sec:r_disc_size}, where the epoch library performs well at small separations from the star. Indeed, the two targets for which the S/N was measured at $\sim$\,20px (0.25\arcsec{}), HD\,1000453 and TW\,Hya, achieved a normalised S/N of 0.95 and 1.00, respectively, when reduced with the epoch library. 

Interestingly, the relative performance of the seeing library compared to the other selection metrics is similar for both the S/N and 20px radius contrast. Given that only one of the disc observations in the sample suffers from wind effects (see Table~\ref{tab:obs_sci}), and following the contrast results for targets with average and bad seeing discussed in Sect.~\ref{sec:r_obs_type}, we could expect that the seeing library would also perform well in terms of S/N. Instead, the bias towards small radii discs has a greater effect on the S/N performance than the observing condition does, suggesting that separation is a more important factor to consider when optimising reference libraries.

Surprisingly, the $\tau_0$ libraries, which yield relatively good contrast results for all but the targets with LWE, give a lower mean S/N than over half of the selection metrics explored. Conversely, reductions using a library of reference targets observed with the same DIT as the science target perform much better in terms of S/N than they do for contrast. This may be explained in part by uncertainties introduced in the measurement process. While the regions in which the S/N was measured were selected to avoid negative signal, this was not possible for every target while also including the brightest disc signal for each reference library reduction and number of PCs. Additionally, some of the brightest areas may not have been included when measuring the S/N of a particular reduction if that area included negative signal for many of the other reductions of the target. We have found that slight variations of the measurement regions can change the ordering of selection metrics with similar mean S/N values. As such, the exact ordering of the selection metrics from highest to lowest mean S/N should not be taken in stone.  

Furthermore, while we favour using the overall distribution of a selection metric to assess its performance, this is a bias arising from the goal of this study: to find a single selection metric to use for a large-scale, blind disc search. Performing multiple reductions with different selection metric libraries may be a better strategy for optimising the S/N of a single disc target. This is highlighted by the M$_\text{H}$ library, which has the second worst performance on average and yet yields the highest S/N for the disc of TWA\,11.

It should also be noted that, since the analysis has been performed on real disc targets, we cannot know how well the morphologies of the discs have been conserved during the reductions, including the ways that complex structures may have been warped or biased by reduction artefacts and over-subtraction. Thus, while the S/N is a useful metric for assessing the reduction capabilities of the different reference libraries, it does not encompass the full impact these libraries have on the disc. \citet{Juillard2024} attempt to address this challenge by performing reductions on cubes with injected fake discs, and using the SSIM to measure the image fidelity, comparing the reduced disc to the ground truth. This analysis is outside of the scope of this paper but could be performed as a future study with our sample of disc-free targets.

\section{Conclusions}
\label{sec:concl}

We have investigated the performance of reference libraries built using various selection metrics on the RDI reduction of synthetic and real discs in order to determine the method to use in the large-scale reduction of archival SPHERE/IRDIS data in the search for new discs. The selection metrics included atmospheric, meteorological, telescope, and stellar parameters, used to preselect reference frames before calculating and then selecting those with the highest frame-to-frame PCC in the speckle-dominated region. 

For the contrast analysis performed on disc-free observation sequences, using synthetic pole-on discs for throughput estimation, we find that the overall best mean normalised contrast is achieved using libraries built using only PCC as a selection metric. The smallest maximum deviation from the best contrast and best S/N performance is achieved using diverse (mixed) reference libraries populated with subsets of frames that closely match different parameters.

Additionally, for the synthetic discs with a 20px radius, reference libraries built using frames observed close in time to the science target observation yield the best contrast for $\sim$\,30\% of the sample. This is supported by the disc S/N analysis, for which the discs with radii at $\sim$\,20px achieve normalised S/N values $\geq$\,0.95 when reduced with the epoch libraries, making epoch a good selection metric for targeted searches of small separation discs.

Grouping science targets by their observing conditions, we find that PCC libraries result in better contrast for targets with LWE, and mixed libraries for targets with WDH. For targets without wind effects, the PCC, seeing, and mixed libraries all give the best mean normalised contrast within $\sim$\,5\%. Seeing, however, may not be an ideal choice of selection metric if the radius of the disc is not known prior to reduction, as it achieves a poorer contrast and S/N performance at smaller separations. Since the seeing library does not yield a high S/N despite all but one of the real disc observations being without wind effects, disc radius may be more important for determining the choice of reference library selection metric than observing condition.

Although the PCC libraries yield the best mean normalised contrast, for reductions with large datasets the computational cost of creating a reference library using only PCC as a selection metric is far greater than for libraries with a preselection step. Since the mixed libraries achieve the best S/N performance, an overall mean normalised contrast within 4\% of the PCC libraries, and the smallest deviation from the best contrast, we prefer them as the choice of selection metric for large-scale data reduction. For the application to individual disc targets, where computational cost is not a concern, we believe that performing multiple RDI reductions using the different selection metric libraries is the best strategy for optimising the reduction of a given observation. 

Alongside the future reduction of SPHERE/IRDIS data, this method could be applied to the reduction of data from other ground-based instruments such as the Gemini Planet Imager, which, similar to SPHERE, benefits from a large set of archival observations and atmospheric measurements.

\begin{acknowledgements}

This project received funding from the European Research Council (ERC) under the European Union's Horizon 2020 research and innovation programme (COBREX; grant agreement n° 885593).

This work has made use of the High Contrast Data Centre, jointly operated by OSUG/IPAG (Grenoble), PYTHEAS/LAM/CeSAM (Marseille), OCA/Lagrange (Nice), Observatoire de Paris/LIRA (Paris), and Observatoire de Lyon/CRAL, and is supported by a grant from Labex OSUG@2020 (Investissements d'avenir - ANR10 LABX56).

This work is based on observations made with ESO Telescopes at the Paranal Observatory.

This research has made use of the SIMBAD database, operated at CDS, Strasbourg, France.

\\
In addition to the software already cited in the paper, we acknowledge the use of the \texttt{SAOImage DS9} visual display utility \citep{SAO2000, Joye2003} and the following Python packages: NumPy \citep{Harris2020_numpy}, SciPy \citep{Virtanen2020_scipy}; Matplotlib \citep{Hunter2007_matplotlib}, OpenCV \citep{Bradski2008_opencv}; Photutils \citep{Bradley2022_photutils}, and Pandas \citep{Pandas2022, McKinney2010_pandas}.

\end{acknowledgements}

\bibliographystyle{aa}
\bibliography{rdi_bibliography}

\newpage
\onecolumn
\begin{appendix}

\begin{landscape}
\section{Science target metadata}
\label{apx:obs_sci}

\begin{table}[h]
    \centering
    \begin{threeparttable}
        \caption{Science target observation sequences used in this paper and their metadata.}
        \label{tab:apx_obs_sci_full}
        \begin{tabular}{l|l|l|l|l|l|l|l|l|l|l|l|l|l|l|l}
            \hline
            Target name & SpT & M$_\mathrm{G}$ & M$_\mathrm{H}$ & Observation night & Condition & Filter & DIT & Frames & $\Delta$PA & Elev. & Seeing & $\tau_0$ & 30m wind & \multicolumn{2}{c}{200mbar wind} \\
             &  &  &  & YYYY-MM-DD &  &  & (s) &  & ($^\circ$) & ($^\circ$) & (\arcsec{}) & (m/s) & speed (m/s) & speed (m/s) & dir. ($^\circ$) \\
            \hline
            BD-19 4288 & G9IV(e) & 10.08 & 7.81 & 2015-04-28 & LWE & H23 & 32 & 48 & 70.5 & 85.2 & 0.92 & 3.34 & 2.23 & 25.15 & 282.6 \\
            HD 9054 & K1V & 8.98 & 6.94 & 2016-11-16 & Bad seeing & H23 & 32 & 160 & 41.8 & 61.8 & 1.91 & 1.87 & 12.65 & 39.26 & 279.2 \\
            HD 13724 & G3/5V & 7.74 & 6.48 & 2018-08-18 & Good seeing & H23 & 32 & 256 & 70.4 & 65.7 & 0.39 & 8.14 & 3.58 & 38.23 & 269.5 \\
            HD 16978 & B9Va & 4.11 & 4.43 & 2016-09-15 & Good seeing & H23 & 32 & 160 & 29.1 & 46.3 & 0.43 & 9.01 & 8.05 & 16.71 & 229.0 \\
            HD 37306 & A1V & 6.09 & 5.99 & 2015-10-28 & Average & H23 & 32 & 48 & 29.2 & 77.0 & 0.97 & 1.80 & 1.70 & 46.70 & 308.9 \\
            HD 37484 & F4V & 7.17 & 6.29 & 2017-12-01 & Bad seeing & H23 & 32 & 144 & 135.5 & 84.0 & 2.61 & 2.60 & 17.73 & 35.21 & 318.8 \\
            HD 59967 & G3V & 6.50 & 5.25 & 2018-03-27 & Average & H23 & 32 & 96 & 50.8 & 76.9 & 0.59 & 9.78 & 5.22 & 12.04 & 216.2 \\
            HD 69830 & G8V & 5.76 & 4.36 & 2018-12-18 & Average & H23 & 32 & 96 & 37.3 & 75.3 & 0.61 & 6.09 & 7.35 & 35.80 & 313.8 \\
            HD 107301 & B9V & 6.20 & 6.28 & 2016-01-18 & Bad seeing & H23 & 32 & 128 & 20.5 & 42.3 & 1.85 & 1.77 & 9.00 & 29.85 & 293.9 \\
            HD 113902 & B8/9V & 5.69 & 5.75 & 2016-06-05 & WDH & H23 & 32 & 80 & 20.4 & 60.7 & 0.93 & 2.05 & 13.79 & 68.69 & 270.9 \\
            HD 121156 & K2III & 5.74 & 3.70 & 2017-02-05 & Good seeing & H23 & 32 & 30 & 60.4 & 85.7 & 0.55 & 6.20 & 10.07 & 27.55 & 271.4 \\
            HD 123247 & B9.5V & 6.42 & 6.47 & 2015-04-05 & LWE & H23 & 32 & 48 & 17.5 & 65.8 & 1.38 & 1.51 & 1.33 & 35.53 & 303.3 \\
            HD 126062 & A1V & 7.45 & 7.43 & 2016-04-08 & Average & H23 & 32 & 128 & 40.7 & 67.1 & 0.54 & 4.93 & 8.45 & 34.21 & 315.7 \\
             & & & & 2016-07-23 & WDH & H23 & 32 & 80 & 25.2 & 66.9 & 1.02 & 2.14 & 8.16 & 48.44 & 284.6 \\
            HD 126135 & B8V & 6.95 & 7.07 & 2018-03-16 & Good seeing & H23 & 32 & 64 & 26.2 & 73.0 & 0.53 & 8.61 & 5.25 & 16.71 & 296.7 \\
            HD 128987 & G8Vk & 7.06 & 5.63 & 2016-05-31 & WDH & H23 & 32 & 96 & 69.4 & 81.1 & 0.89 & 2.25 & 9.25 & 56.77 & 269.3 \\
            HD 147808 & G9IVe & 9.34 & 7.28 & 2016-04-14 & LWE & H23 & 32 & 48 & 6.7 & 80.4 & 0.31 & 7.79 & 4.10 & 28.16 & 286.3 \\
            HD 165185 & G1V & 5.81 & 4.61 & 2019-09-06 & WDH & H23 & 32 & 128 & 64.6 & 77.4 & 1.26 & 1.79 & 16.00 & 50.24 & 298.8 \\
            HD 188228 & A0Va & 3.96 & 3.76 & 2015-06-30 & LWE & H23 & 32 & 128 & 21.3 & 41.2 & 1.07 & 1.20 & 0.55 & 61.20 & 266.1 \\
            HD 213398 & A1Va & 4.26 & 4.30 & 2016-06-13 & Bad seeing & H23 & 32 & 208 & 121.2 & 80.1 & 1.68 & 1.57 & 14.25 & 38.56 & 263.6 \\
            \hline
            2MASS J1604$^{(\text{a})}$ & K2 & 11.71 & 9.10 & 2015-06-09 & Bad seeing & K12 & 32 & 176 & 142.9 & 82.8 & 1.39 & 2.10 & 11.05 & 21.60 & 295.0 \\
            HD 100453 & A9Ve & 7.74 & 6.39 & 2016-01-19 & Bad seeing & H23 & 64 & 64 & 31.3 & 60.0 & 1.56 & 2.30 & 9.21 & 27.60 & 284.2 \\
            HD 111520 & F5/6V & 8.83 & 7.83 & 2016-05-14 & WDH & H23 & 32 & 128 & 36.2 & 64.4 & 1.17 & 1.82 & 13.62 & 41.69 & 264.2 \\
            HD 141569 & A2/B9V & 7.10 & 6.86 & 2015-05-15 & Average & H23 & 64 & 64 & 42.1 & 68.9 & 0.81 & 3.60 & 10.66 & 8.91 & 276.2 \\
            HD 181327 & F6V & 6.94 & 5.98 & 2015-05-09 & Average & H23 & 64 & 56 & 31.1 & 60.0 & 1.34 & 1.71 & 4.50 & 34.91 & 228.3 \\
            HR 4796 & A0V & 5.78 & 5.79 & 2015-02-02 & Average & H23 & 32 & 112 & 48.5 & 74.2 & 0.67 & 11.68 & 2.58 & 11.60 & 350.3 \\
            TW HYA & K6Ve & 10.45 & 7.56 & 2015-02-03 & Average & H23 & 64 & 64 & 76.7 & 79.1 & 0.57 & 11.30 & 4.70 & 7.84 & 331.5 \\
            \hline
        \end{tabular}
        \tablefoot{
            Disc-free targets are listed above the horizontal line, and disc targets below the line. $\Delta$PA refers to the change in parallactic angle during the observation sequence. Other parameters are defined as in Table~\ref{tab:params}, and take the median value of the full sequence where applicable.\\
            \tablefoottext{a}{2MASS J16042165-2130284.}
        }
    \end{threeparttable}
\end{table}
\end{landscape}

\section{Disc signal region}
\label{apx:disc_reg}

\begin{figure}[h]
    \centering
    \includegraphics[width=18.4cm]{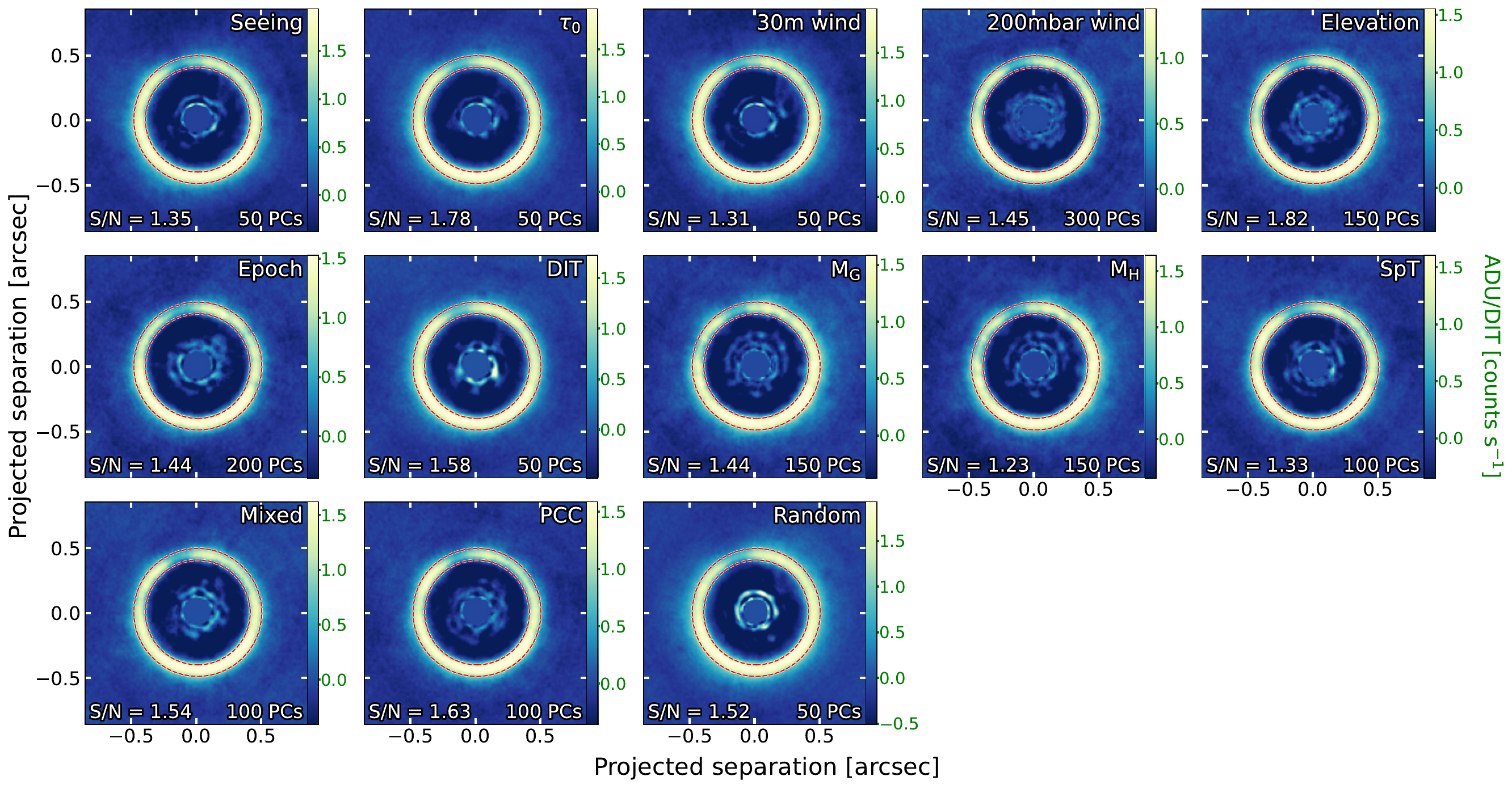}
    \caption{2MASS J16042165-2130284 2015-06-09 IRDIS-K12 RDI-PCA reductions using the different reference libraries. The reduction with the best S/N is shown for each library. Each image is labelled with the selection metric of the reference library (top right), the number of PCs used (bottom right), and the mean S/N of the disc (bottom left). The region in which the S/N was measured is indicated by the red dashed lines.}
    \label{fig:J1604}
\end{figure}

\begin{figure}[h]
    \centering
    \includegraphics[width=18.4cm]{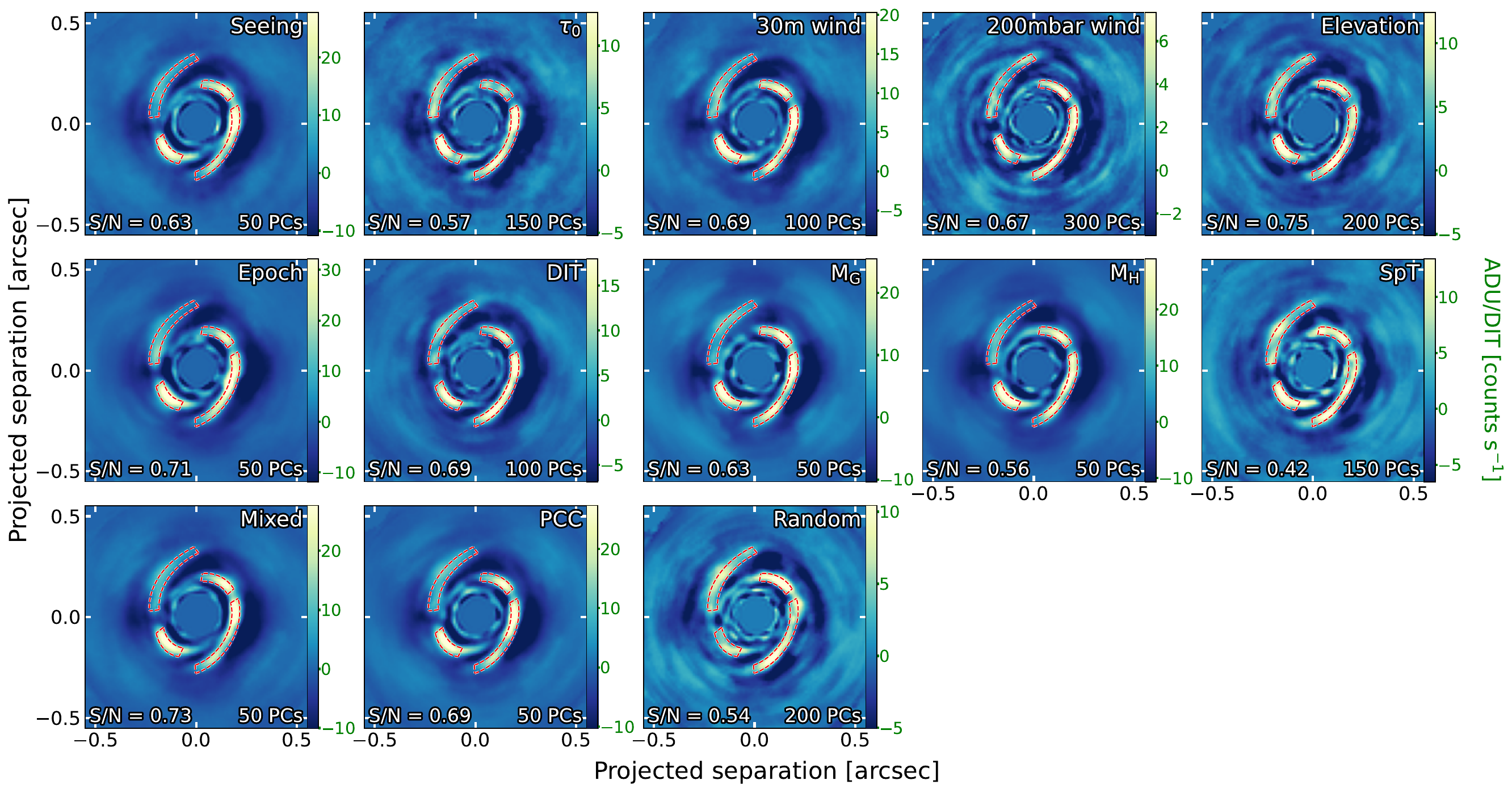}
    \caption{HD 100453 2016-01-19 IRDIS-H23 RDI-PCA reductions. Layout as described in the caption of Fig.~\ref{fig:J1604}.}
    \label{fig:HD100453}
\end{figure}

\begin{figure}
    \centering
    \includegraphics[width=18.4cm]{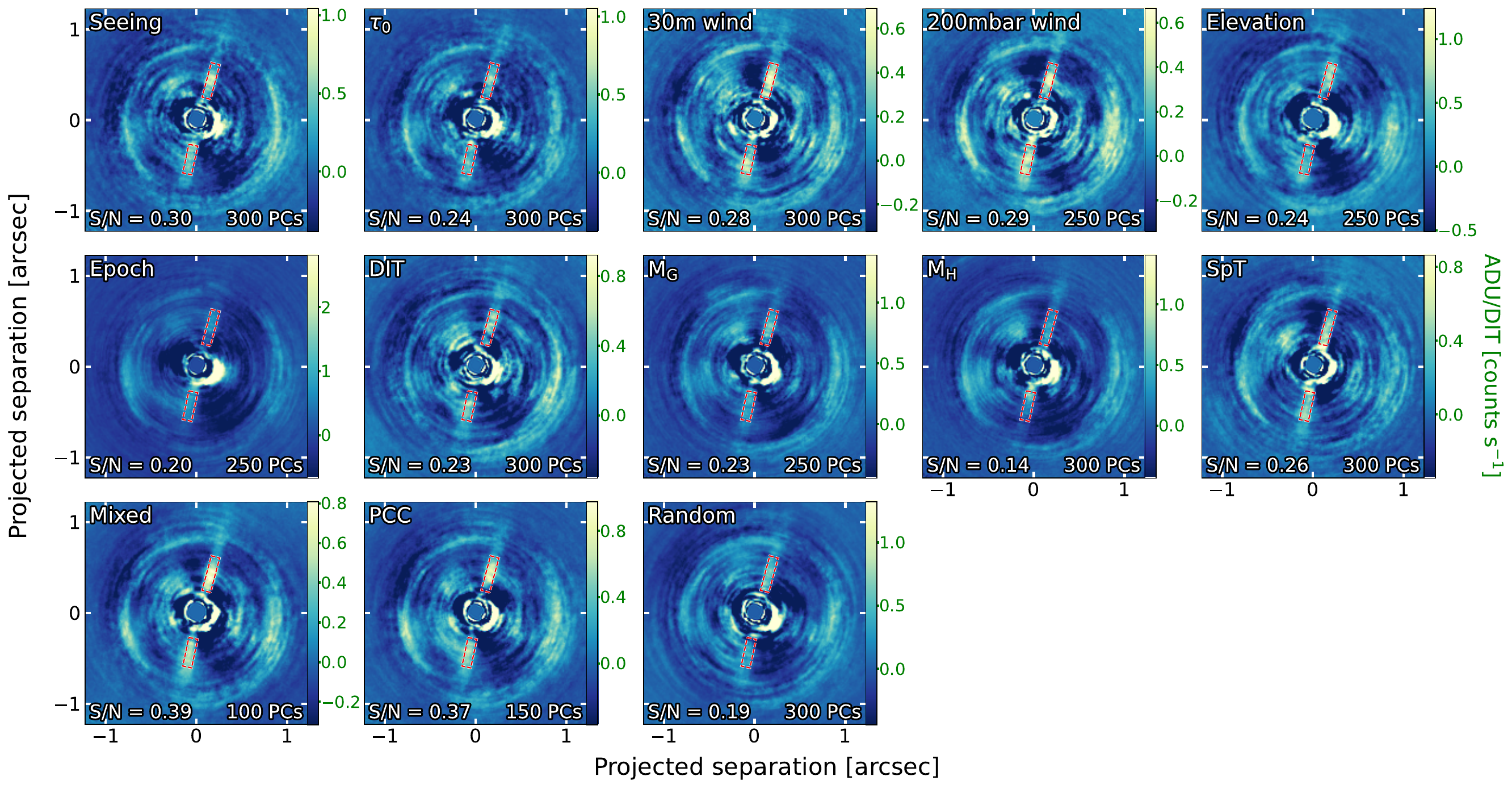}
    \caption{HD 111520 2016-05-14 IRDIS-H23 RDI-PCA reductions. Layout as described in the caption of Fig.~\ref{fig:J1604}. The selection metric is labelled in the top left of each image.}
    \label{fig:HD111520}
\end{figure}

\begin{figure}
    \centering
    \vspace{15pt}
    \includegraphics[width=18.4cm]{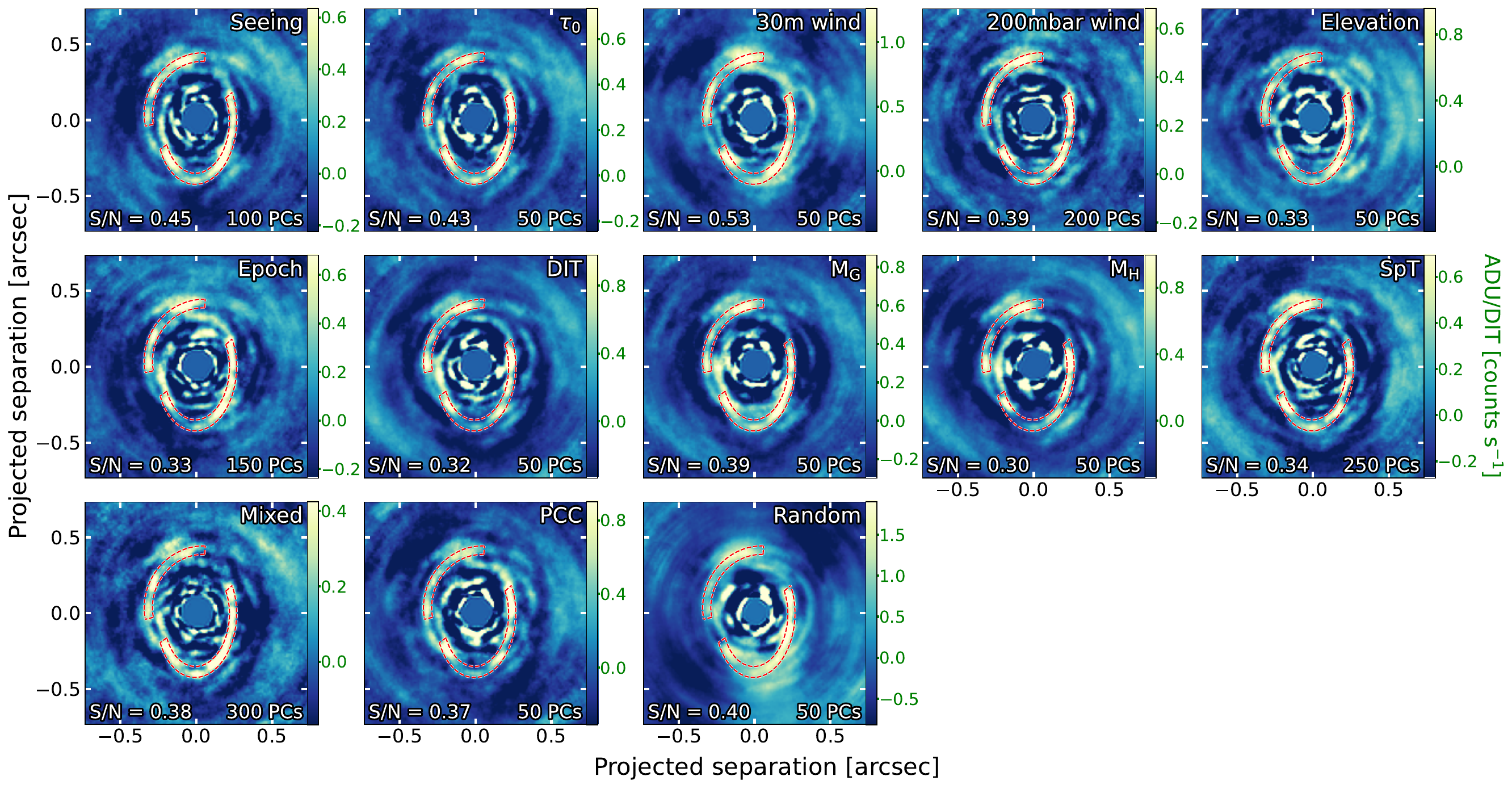}
    \caption{HD 141569 2015-05-15 IRDIS-H23 RDI-PCA reductions. Layout as described in the caption of Fig.~\ref{fig:J1604}.}
    \label{fig:HD141569}
\end{figure}

\begin{figure}
    \centering
    \includegraphics[width=18.4cm]{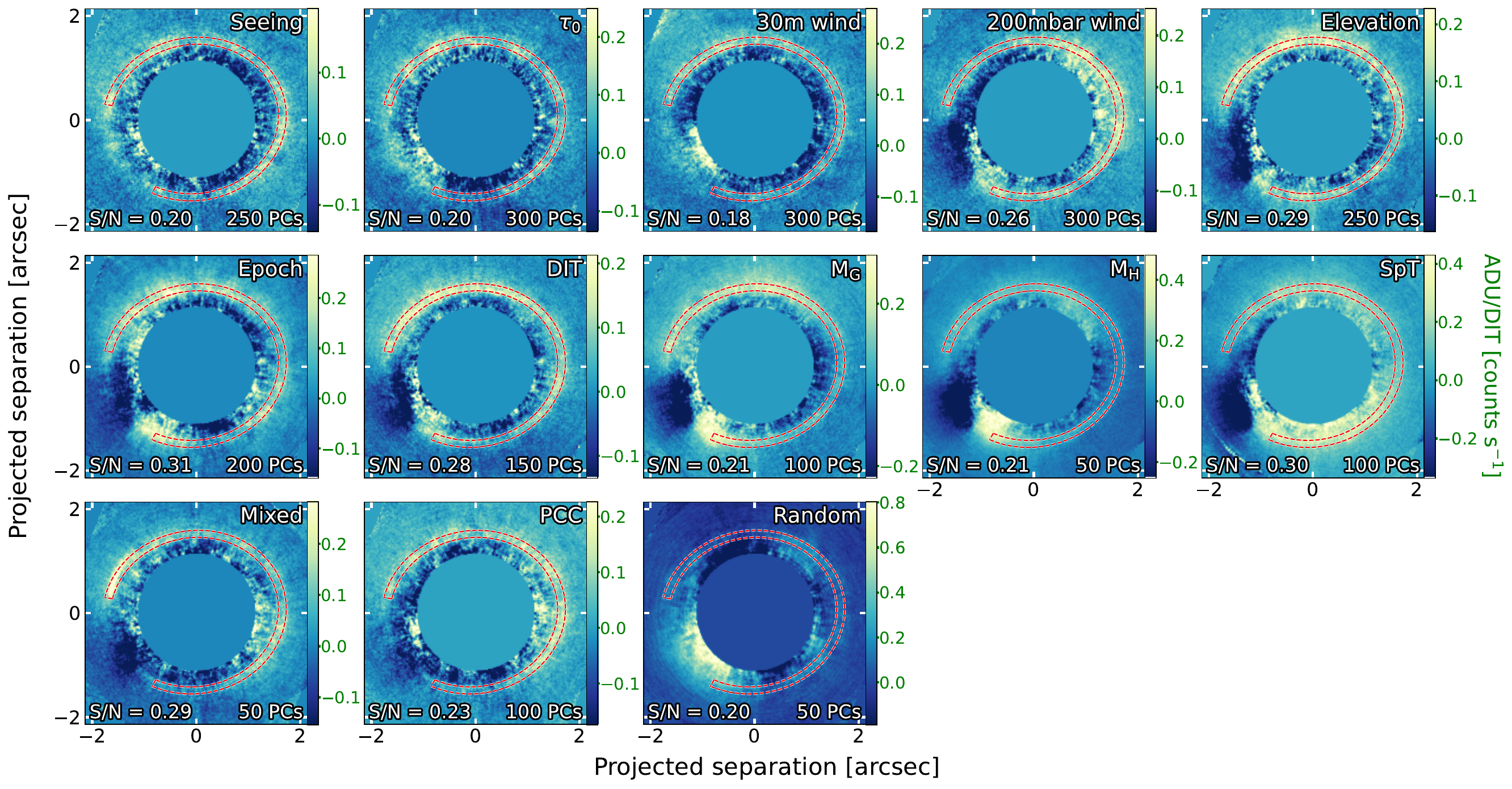}
    \caption{HD 181327 2015-05-09 IRDIS-H23 RDI-PCA reductions. Layout as described in the caption of Fig.~\ref{fig:J1604}.}
    \label{fig:HD181327}
\end{figure}

\begin{figure}
    \centering
    \vspace{15pt}
    \includegraphics[width=18.4cm]{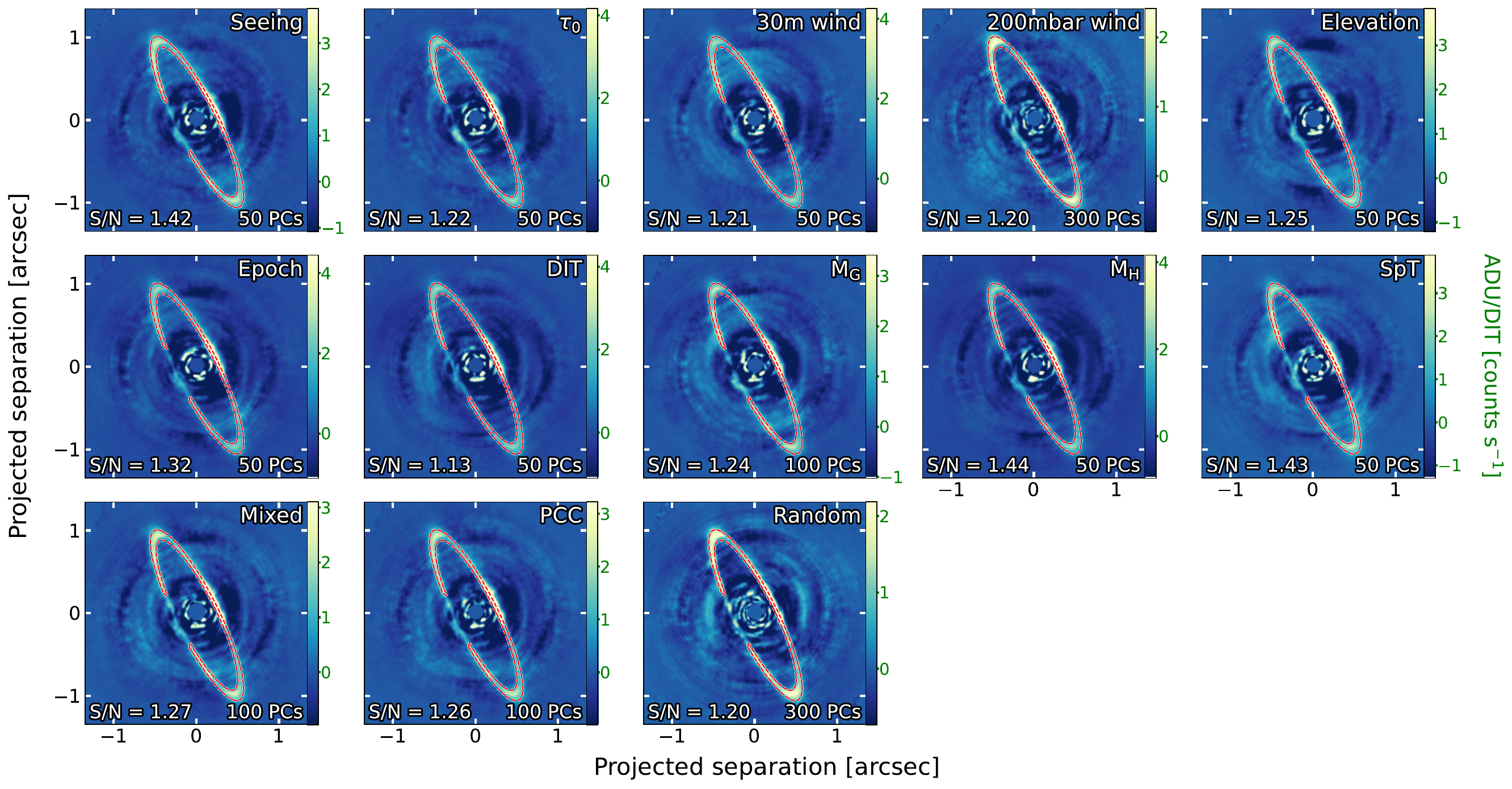}
    \caption{HR 4796 2015-02-02 IRDIS-H23 RDI-PCA reductions. Layout as described in the caption of Fig.~\ref{fig:J1604}.}
    \label{fig:HR4796}
\end{figure}

\begin{figure}
    \centering
    \includegraphics[width=18.4cm]{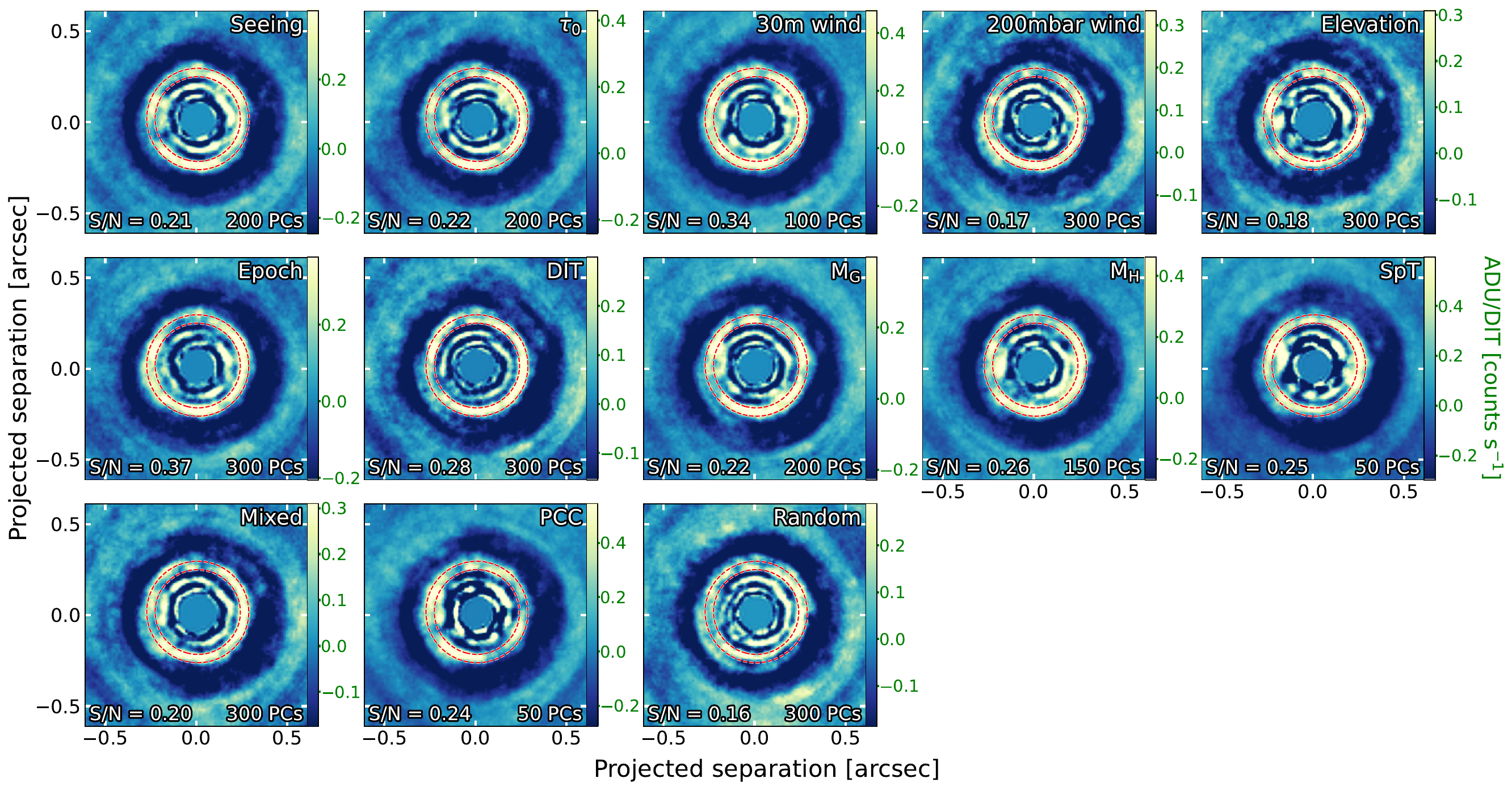}
    \caption{TW HYA 2015-02-03 IRDIS-H23 RDI-PCA reductions. Layout as described in the caption of Fig.~\ref{fig:J1604}.}
    \label{fig:TWHYA}
\end{figure}

\end{appendix}

\end{document}